\documentclass[11pt,twoside]{article}

\usepackage{baaa2013}

\usepackage{graphicx}
\usepackage{subfigure}
\usepackage{amssymb}
\usepackage[spanish,activeacute,english]{babel}
\usepackage[latin1]{inputenc}
\usepackage{verbatim}
\usepackage{amsmath}
\usepackage{amsfonts}
\usepackage{amssymb}
\usepackage[colorlinks=true,dvips]{hyperref}
\usepackage{natbib}

\begin{document}
%%%%%%%%%%%%%%%%%%%%%%%%%%%%%%%%%%%%%%%%%%%%%%%%%%%%%%%%%%%%%%%%%%%%%%%%%%
%%%% SELECCIONE EL IDIOMA EN QUE SE ESCRIBE EL ARTÍCULO:              %%%%
%\myselectspanish
\myselectenglish
%%%%%%%%%%%%%%%%%%%%%%%%%%%%%%%%%%%%%%%%%%%%%%%%%%%%%%%%%%%%%%%%%%%%%%%%%%
\vskip 1.0cm
\markboth{De Rossi et al.}%
{The baryonic mass assembly of low-mass halos in a $\Lambda$-CDM Universe}

\pagestyle{myheadings}
\vspace*{0.5cm}

%%%% INDIQUE LA CATEGORIA QUE DESCRIBE EL CARACTER DE SU TRABAJO       %%%%
%1 = PRESENTACION MURAL
%2 = PRESENTACIÓN ORAL
%3 = TRABAJO INVITADO 
%4 = PRESENTACION EN MESA REDONDA
%5 = PREMIO

\Categ{2}

%%%% INDIQUE EL TEMA AL QUE CORRESPONDE SU TRABAJO       %%%%
%1 = Sol, Sistema Solar y Ciencias Planetarias
%2 = Estrellas y Sistemas Estelares
%3 = Medio Interestelar
%4 = Estructura Galáctica
%5 = Astronomía Extragaláctica
%6 = Cosmología
%7 = Astrofísica de Altas Energías
%8 = Instrumentación y Sitios Astronómicos
%9 = Otros
%10 = Divulgación, Enseñanza e Historia

\Tema{5}

\vskip 0.3cm
\title{The baryonic mass assembly of low-mass halos in a $\Lambda$-CDM Universe}

%\title{ Template paper for publication in the Bulletin of the 
%Argentinian Astronomical Association with instructions for the use of 
%\LaTeX{}}
 
\author{M.E. De Rossi$^{1,2,3}$, V. \'Avila{\--}Reese$^{4}$, P.B. Tissera$^{1,2, 5}$, A. Gonz\'alez-Samaniego$^{4}$ \& S.E. Pedrosa$^{1,2}$}

\affil{%
  (1) Instituto de Astronomía y Física del Espacio (CONICET-UBA)\\ 
  (2) Consejo Nacional de Investigaciones Cient\'ificas y T\'ecnicas, CONICET, Argentina (derossi@iafe.uba.ar)\\ 
  (3) Departamento de Física, Facultad de Ciencias Exactas y Naturales, Universidad de Buenos Aires, Argentina\\
  (4) Instituto de Astronom\'ia, Universidad Nacional Aut\'onoma de M\'exico, A.P- 70-264, 04350 M\'exico, D.F., M\'exico \\
  (5) Departamento de Ciencias Físicas, Universidad Andres Bello, Av. Republica 220, Santiago, Chile (current address)
}

\begin{abstract} 
We analyse the dark, gas, and stellar mass assembly histories of low-mass
halos ($M_{\rm vir} \sim 10^{10.3} \-- 10^{12.3} \ M_{\odot}$)
identified at redshift $z=0$
in cosmological numerical simulations. 
Our results indicate that  for halos in a given present-day mass bin,
the gas-to-baryon fraction inside the virial radius
does not evolve significantly with time, ranging
from $\sim 0.8$ for smaller halos to $\sim 0.5$ for the largest ones.
Most of the baryons are located actually not in the
galaxies but in the intrahalo gas; for the more massive halos,
the intrahalo gas-to-galaxy mass ratio is approximately
the same at all redshifts, $z$, but for the least massive halos, it
strongly increases with $z$. The intrahalo gas in the former halos gets hotter
with time, being dominant at $z=0$, while in the latter halos, it is 
mostly cold at all epochs. The multiphase ISM and thermal feedback
models in our simulations work in the direction of 
delaying the stellar mass growth of low-mass galaxies.
\end{abstract}

\begin{resumen}
Analizamos las historias de ensamblaje de la masa oscura, gaseosa y estelar 
de halos de baja masa ($M_{\rm vir} \sim 10^{10.3} \-- 10^{12.3} \ M_{\odot}$) 
identificados a corrimiento al rojo $z=0$ en simulaciones numéricas cosmológicas.
Nuestros resultados indican que, para halos en un dado rango actual de masa,
la fracción de gas respecto de bariones dentro del radio virial no evoluciona
significativamente con el tiempo,  extendiéndose desde $\sim 0.8$ para los
halos pequeños hasta $\sim 0.5$ para los más masivos.
En realidad, la mayor parte de los bariones no está localizada en las galaxias
sino en el gas intra-halo;  para los halos más masivos, el cociente de masas
entre el gas intra-halo y la galaxia central es aproximadamente el mismo a 
todo corrimiento al rojo, $z$, pero, para los menos masivos, éste se 
incrementa fuertemente con $z$.  En los primeros, el gas intra-halo 
aumenta su temperatura con el tiempo, siendo la componente caliente dominante
a $z=0$, mientras que en los segundos, la componente gaseosa es predominantemente 
fría en toda época.  En nuestras simulaciones, los modelos de ISM multi-fase y {\it feedback} térmico
favorecen el retraso del crecimiento de la masa estelar
en galaxias de baja masa.
\end{resumen}

\section{Introduction}
\label{S_intro}
Cosmological simulations of structure formation constitute fundamental tools
for studying the problem of galaxy evolution through cosmic epochs.
In the $\Lambda$ cold
dark matter (CDM) paradigm, more massive halos assemble by the hierarchical aggregation
of smaller ones (upsizing).  Galaxies form from the gas trapped inside the potential well of 
these halos and are affected, in a complex way, by different astrophysical processes such as 
gas cooling, star formation, supernova (SN) feedback, and chemical enrichment, among others.
In this context, a current challenge for observational and theoretical studies 
is the understanding of the relation between the halo mass aggregation histories and the 
evolution of the baryonic matter hosted by these halos, both in galaxies and in the
intergalactic medium around them.

The relation between the stellar ($M_*$) and virial ($M_{\rm vir}$) mass of galaxies
has been the subject of many empirical and semi-empirical studies in the local 
Universe and at higher redshifts (see e.g., Firmani \& Avila-Reese 2010,
Avila-Reese \& Firmani 2011).
From a theoretical point of view, current models and simulations
of low-mass galaxies evolving in a $\Lambda$-CDM universe seem to predict
a too early $M_*$ assembly than what observations suggest (Avila-Reese et al. 2011 and more
references therein).
In De Rossi et al. (2013; DR+13), we analysed the halo, baryonic and stellar mass assembly
of galaxies by means of cosmological numerical simulations.
We have found that the upsizing trend
of halo mass growth seems to be reverted to a moderately downsizing trend
in the case of the galaxy baryonic/stellar mass assembly, though
the latter trend is weaker than the observed one.
Here, we present results from these simulations regarding the
evolution of the intrahalo gas, which is affected by the halo capture, heating and cooling
mechanisms as well as by the galaxy SN-driven outflows.

\section{Simulations}

We performed numerical simulations by using the chemical code GADGET-3 (Scannapieco et al. 2008),
which includes treatments for metal-dependent radiative cooling, stochastic star formation,
chemical enrichment and  SN feedback.
We assumed the concordance $\Lambda$-CDM cosmology
with $\Omega_m =0.3, \Omega_\Lambda =0.7, \Omega_{b} =0.045$, a normalisation of the power
spectrum of ${\sigma}_{8} = 0.9$ and $H_{0} =100 \, h$ km s$^{-1} \ {\rm Mpc}^{-1}$ with  $h=0.7$.
The simulated volume corresponds to a cubic box of a comoving 10 Mpc $h^{-1}$ side length.
The masses of dark matter and initial gas-phase particles are
$5.93 \times 10^6 \ {\rm M}_{\sun} h^{-1}$ and  $9.12 \times 10^5 \ {\rm M}_{\sun} h^{-1}$, respectively (S230) 
and $2.20 \times 10^6 \ {\rm M}_{\sun} h^{-1}$ and  $3.40 \times 10^5 \ {\rm M}_{\sun} h^{-1}$, respectively (S320).
%Due to computational costs, the box S320 was run only up to z=2.
See De Rossi et al. (2010, 2012) and DR+13 for more details about the simulations.

\section{Results}

We analyse the mass aggregation histories (MAHs) associated to halos
identified at $z=0$ by separating them into four mass bins accordingly to their  
$\log ( M_{\rm vir} / {\rm M}_{\sun} ): <10.5 , 10.5 - 11.0, 11.0-11.5, \ge 11.5$.  
As discussed in detail in DR+13, 
the average MAHs are in rough agreement with those derived from the 
Millennium Simulations, though for the lower-mass halos, at high $z'$s, our
halos are systematically less massive. The latter is due to the effects of baryons
on the total halo mass. 
%for each bin, 
%the simulated $M_{\rm vir}$ increases towards lower redshifts in general
%agreement with the associated mean halo MAH derived from the Millennium simulations.
%Nevertheless, there are minor systematic differences at $z > 1$, specially in the case
%of smaller systems, indicating a later mass assembly in our galaxies. 
In general, both the Millennium
Simulations and our runs show a later $M_{\rm vir}$ assembly as more massive are
the halos (upsizing).  We also analyse the {\it total} stellar and baryonic masses inside the 
virial radius ($R_{\rm vir}$) and find that they assemble approximately at the same 
rate as the halo mass. 
Inside the halos, more than half of the baryons are always in the gas phase 
at all our scales and at least up to $z\sim 3$. This gas-to-baryon mass fraction 
ranges on average from $\sim 0.8$ for low-mass systems to $\sim 0.55$ for the more 
massive ones, with little change with $z$. This means that the lower the halo mass, the less 
efficient is the system in making stars. 

\begin{figure}[!ht]
\centering
\includegraphics[width=0.51\textwidth]{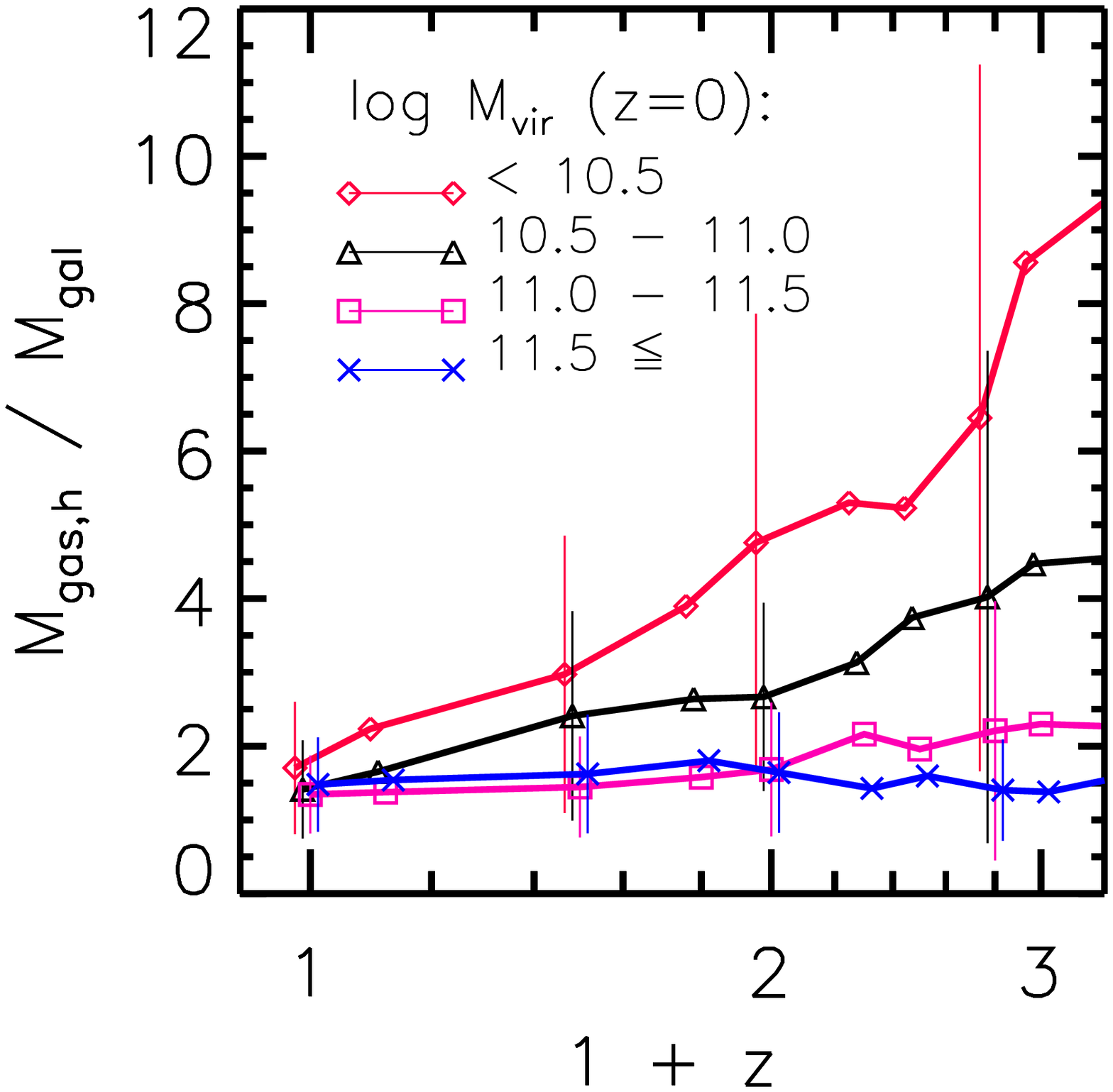} \hspace{-0.55cm}
\includegraphics[width=0.51\textwidth]{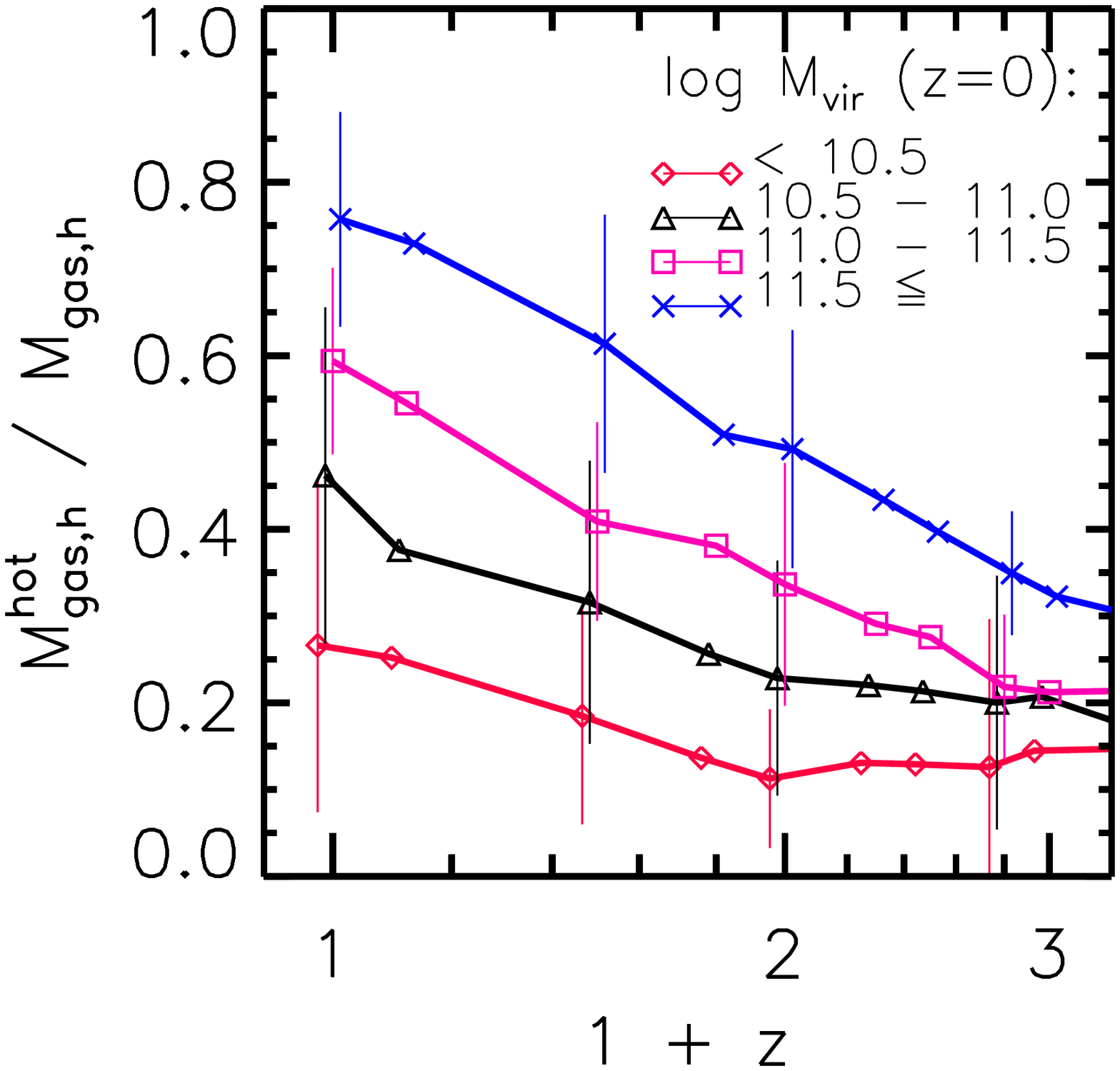}
\caption{
Average intrahalo gas-to-total galaxy mass ratio (left panel) and the fraction of the
intrahalo gas in the warm-hot phase (right panel) for four $\log (M_{\rm vir} / M_{\odot})$ bins 
defined at $z = 0$ (solid lines):
$<10.5$ (red diamonds), $10.5-11$ (black triangles), $11-11.5$ (pink squares) and $\ge 11.5$ (blue crosses).
Error bars represent $1\sigma$ population scatter.  
}
\label{fig:fig1}
\end{figure}

In Fig. \ref{fig:fig1}, left panel, we show the evolution of the ratio of the gas mass in the halo
(i.e. outside the galaxy but inside $R_{\rm vir}$), $M_{\rm gas,h}$, to the galaxy
(stars+gas) mass, $M_{\rm gal}$, for the four aforementioned mass bins. 
At $z\sim 0$, the mass in the intrahalo gas is on average $\sim 1.5\times$ larger than
the mass in the central galaxy (satellites are typically very small). This ratio
is roughly the same at higher $z'$s for the massive halos, but it significantly
increases with $z$ for the less massive halos. 
In the right panel of Fig. \ref{fig:fig1}, we analyze the evolution of the fraction of this 
intrahalo gas in warm-hot and cold phases, with a temperature separation of $T = 15000 \ K$.
For the lowest-mass halos, the intrahalo gas is mostly cold ($\sim 80\%$) at all $z'$s, while for
the most massive halos, the fraction of warm-hot gas increases with time, from
$\approx 30\%$ at $z\sim 3$ to $75\%$ at $z\sim 0$.

\section{Discussion}

Our cosmological simulation of $\sim 300$ low-mass galaxies in a 14.3 Mpc-size box
shows that  the smaller the halo mass, the larger is the total gas-to-baryon mass fraction inside 
$R_{\rm vir}$. Therefore, the efficiency in making stars decreases 
with $M_{\rm vir}$; for $M_{\rm vir}\approx 1-3\times 10^{10} M_{\odot}$, 
more than 80\% of the baryons are in the gas (mostly cold) phase at all $z'$s.  The multiphase ISM
and the thermal feedback models introduced in the code seem to work in the direction
of avoiding too early and active SF in the low-mass systems. 

The simulation shows that most of the baryons in the halos at $z\sim 0$ are located 
actually in the intrahalo gas, which  on average accounts for $\sim 1.5\times$ more 
mass than in the galaxies (left panel, Fig. \ref{fig:fig1}). This suggests that a significant 
fraction of the present-day missing baryons are actually located in gas around the galaxies. 
What is the thermal state of this gas?  This depends strongly on the halo mass
(right panel, Fig. \ref{fig:fig1}): for the smaller systems, most of the gas is in the cold 
phase ($T < 15000$K), while for the larger systems, it dominates the warm-hot 
phase (up to $\sim 75\%$ on average in halos of $M_{\rm vir}\approx 3-30\times 10^{11} M_{\odot}$).  
For the latter, the intrahalo gas significantly warms up with time, as the halo grows and virializes,
limiting this the ulterior growth of the galaxy. For the less massive systems,
the intrahalo gas is cold at all redshifts and it efficiently inflows to the galaxy (cold flows)
in such a way that the intrahalo gas-to-galaxy mass ratio strongly decreases
with time (left panel, Fig. \ref{fig:fig1}).

\acknowledgments 
We acknowledge a CONACyT-CONICET (M\'exico-Argentina) bilateral grant for partial funding.
V.A. and A. G. acknowledge PAPIIT-UNAM grant IN114509.  A.G. acknowledges a PhD fellowship
provided by CONACyT.
M.E.D.R., P.B.T. and S.P. acknowledge support from
PICT 245-Max Planck (2006) of ANCyT (Argentina), 
PIP 2009-112-200901-00305 of CONICET (Argentina), 
PICT Raices (2011) of ANPCyT (Argentina) and 
the L'oreal-Unesco-Conicet 2010 Prize.
Simulations were run in Fenix and HOPE clusters at IAFE and Cecar cluster at University
of Buenos Aires, Argentina.

\begin{referencias}

\reference Avila-Reese, V., \& Firmani, C.\ 2011, Revista Mexicana de Astronomia y Astrofisica Conference Series, 40, 27

\reference Avila-Reese, V. et al. \ 2011, \apj, 736, 134

\reference de Rossi, M.~E., Tissera, P.~B., \& Pedrosa, S.~E.\ 2010, \aap, 519, A89

\reference De Rossi, M.~E., Tissera, P.~B., \& Pedrosa, S.~E.\ 2012, \aap, 546, A52

\reference De Rossi, M.~E., Avila-Reese, V., Tissera, P.~B., A., Gonz\'alez-Samaniego \& Pedrosa, S.~E.\ 2013, \mnras, 435, 2736
(DR+13)

\reference Firmani C. \& Avila-Reese V., 2010, ApJ, 723, 755

\reference Gonz{\'a}lez-Samaniego, A. \& Avila-Reese, V. \ 2013, AAABS, 4, 83

\reference Scannapieco, C., Tissera, P.~B., White, S.~D.~M., \& Springel, V.\ 2008, \mnras, 389, 1137

\end{referencias}

\end{document}